# Perspective: Mitigation of structural defects during the growth of two-dimensional van der Waals chalcogenides by molecular beam epitaxy

Qihua Zhang[1,†], Maria Hilse[1,3], and Stephanie Law[1, 2,3,*]

[1]Two Dimensional Crystal Consortium Materials Innovation Platform, The Pennsylvania State University, University Park, PA 16802, USA

[2]Materials Research Institute, The Pennsylvania State University, University Park, Pennsylvania 16802 USA

[3]Department of Materials Science and Engineering, The Pennsylvania State University, University Park, Pennsylvania 16802 USA

[†]qzz5173@psu.edu

[*]sal6149@psu.edu

**Abstract:** The growth of wafer-scale van der Waals (vdW) thin films and heterostructures by molecular beam epitaxy (MBE) is important for future applications in quantum technologies, next generation optoelectronic devices, and fundamental physics investigations. When grown using co-deposition methods that are typically used for compound semiconductor MBE, vdW materials typically show a high density of structural defects including twin or antiphase domains, spiral growth, and pyramidal growth. These defects are caused by the relatively weak film/substrate interaction and/or the poor wettability of typical substrates by many vdW materials. These difficulties can be mitigated using a multi-step growth procedure in which growth stages including nucleation and coalescence can be rigorously controlled, resulting in high-quality deposition of vdW thin films. This article will describe a general recipe for the growth of highly-crystalline wafer-scale vdW thin films by MBE.

## 1. Introduction

Van der Waals (vdW) materials are materials that have strong bonding within a single layer or unit cell and weak vdW bonding between layers. Due to the weak interlayer bonding, it is possible to stack multiple different vdW materials into one device, either through exfoliation from a single crystal and subsequent stacking or through direct vdW heteroepitaxial growth.[1–5] Although exfoliation and transfer of vdW flakes has proven extremely useful for understanding the fundamental physics of vdW materials and heterostructures and for creating prototype devices, the synthesis of wafer-scale thin films of vdW materials and heterostructures is necessary for scalable device applications. In conventional covalently-bonded materials, epitaxial growth of thin films is one of the most common ways to achieve wafer-scale devices. However, the lattice constant and crystal structure of these materials needs to be similar to the substrate to achieve high-quality layers, which places significant restrictions on which substrates can be used and which materials can be stacked.[6–8] One major benefit of vdW materials is that they can be grown on a wide range of substrates since the film/substrate interaction is relatively weak.[5,9–11] Wafer-scale crystalline vdW thin films have been grown on substrates with very large lattice mismatches as well as on amorphous and flexible substrates, though higher-quality films are generally grown on substrates with similar lattice constants and symmetries or on substrates with vdW bonding.[12–14] The relatively weak film/substrate interaction makes it straightforward to integrate vdW materials with a range of existing devices. However, this benefit comes with a cost. Because the film-substrate interaction is weak, vdW materials are prone to a wide variety of structural defects including defective layers at the film/substrate interface, twin defects (also called antiphase domains), spiral growth, and pyramidal growth.[5,15–19] Because of the low surface energies intrinsic to vdW materials, secondary nucleation faces the same issues as layer growth progresses.[20–23] These defects are therefore nearly impossible to outgrow by increasing the layer thickness or using buffer layers – a well-established strategy in compound semiconductor synthesis to suppress the propagation of defects from (mismatched) interfaces.[24–26] Consequently, these defects lead to increased electron scattering rates, more nonradiative optical recombination pathways, and increased quantum decoherence, severely limiting device performance across a range of spaces. This perspective will discuss the origin of these structural defects as well as strategies to reduce or eliminate them using molecular beam epitaxy (MBE) growth. Many of these strategies may be applicable to other thin film growth techniques as well.

## 2. Molecular beam epitaxy

This article will focus on understanding and improving the growth of vdW thin films by MBE, and this section will explain the fundamentals of MBE. For further details on MBE growth, see refs. [6,8,21,27,28] In short, MBE is an ultra-high vacuum (UHV) deposition technique in which a film is (ideally) deposited atomic layer by layer; a schematic of an MBE chamber is shown in Figure 1(a). The typical MBE chamber contains high purity elemental sources which are thermally evaporated. For example, to grow a $Bi_2Se_3$ film, elemental bismuth and selenium materials are heated in ceramic crucibles until the bismuth and selenium atoms evaporate. Because MBE is a UHV technique, the atoms travel to the substrate as molecular beams, i.e., the mean free paths of the atoms through which they travel without scattering or undergoing gas phase reactions is much longer than the dimensions of an MBE system. In an MBE system with a typical pressure during growth of $10^{-6}$ Torr, the mean free path is ~5 m, much larger than the MBE system itself.[29] For context, the background pressure of an MBE

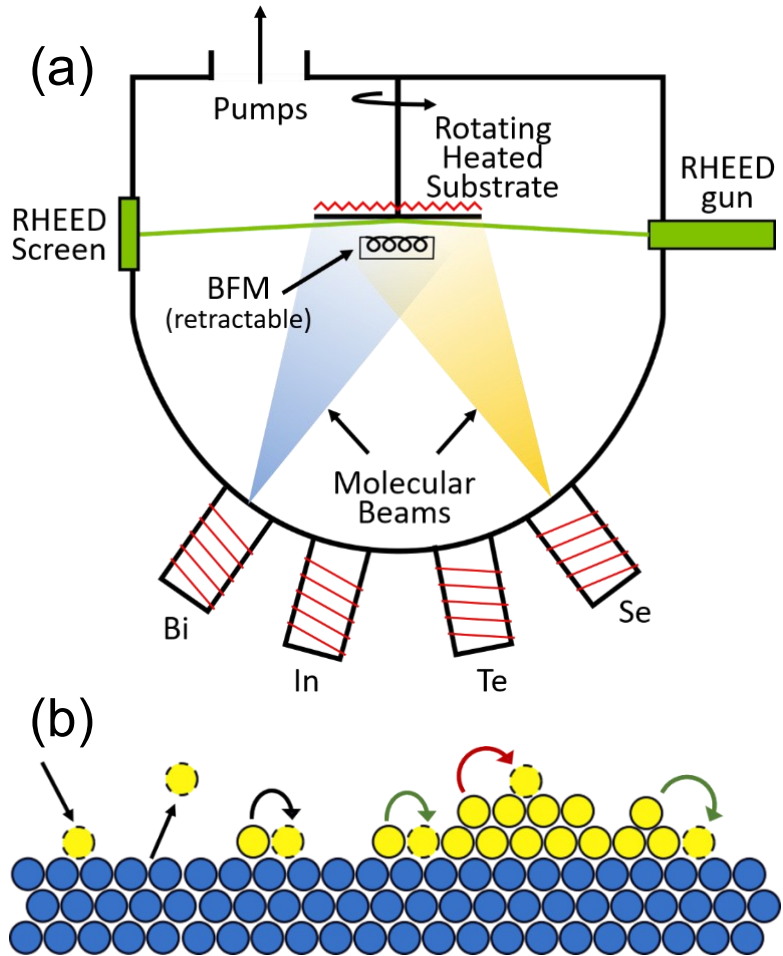

Figure 1 (a) Schematic of an MBE system. (b) Possibilities for adatom motion (yellow circles) on a substrate (blue circles). Adatoms can adsorb (black down arrow), desorb (black up arrow), hop from site to site (black curved arrow), attach to the edge of a domain (green arrows), or hop up from one film layer to the next (red arrow).

system is typically in the range of $10^{-11}$-$10^{-9}$ Torr while idling. The evaporated atoms impinge upon a heated, crystalline substrate and can undergo a variety of kinetic processes with different probabilities, as shown in Figure 1(b), where the yellow circles represent the adatoms comprising the film and the blue circles represent the substrate atoms. The adatoms may reevaporate (black straight up arrow), they may diffuse along the surface from site to site (black curved arrow), they may incorporate at a step edge to start growing the film (green arrows), they may diffuse upward on an existing film island (red arrow), or they may land on an existing film island and incorporate. In ideal layer-by-layer growth, every adatom will either reevaporate or incorporate at a step edge until the layer is complete. For a typical 3D material, incorporation at a step edge is energetically favorable due to the high density of dangling bonds at this location. Under these conditions, an island nucleates and grows laterally through step-edge incorporation until it encounters another island, and they coalesce. This process continues until one complete layer of the film has formed, after which new islands nucleate and grow to form the second atomic layer of the film. This type of growth is typically called step flow growth, and it leads to smooth films with well-defined thicknesses.[30] Conventional semiconductors with covalent bonding can grow in the step flow mode, enabling the creation of extremely complex devices with many layers. However, due to the relatively weak interactions in a vdW film, it is often difficult to establish pure step flow growth, leading to a range of structural defects in epitaxial vdW films, as discussed in the next section.

## 3. Common types of structural defects found in vdW materials

As noted above, uncontrolled growth of vdW materials on 3D substrates often results in films with multiple different types of structural defects. This article is solely focused on epitaxial growth of vdW films on 3D substrates. Many of the structural defects described here are mitigated by growth of vdW films on vdW substrates like graphene or mica. However, growth on vdW substrates is often not appropriate for device applications, and it will not be discussed further. In addition to the structural defects described below, vdW materials can also suffer from point defects, stacking faults, and oxidation. Mitigating these defects can be quite challenging and is beyond the scope of this article. In this section, the origin of the structural defects will be described in detail. Although $Bi_2Se_3$ is the exemplar material, similar issues are found for a wide range of vdW materials. Strategies and prescriptions to mitigate defects will be found in Section 4.

The first common challenge in the growth of vdW materials is obtaining long-range crystalline order in the film at the film/substrate interface. Because the film/substrate interaction is weak, vdW materials can be grown on a wide range of substrates. However, this weak interaction means that there is no strong energetic driver to force the adatoms to nucleate with high crystallinity and long-range order. This is especially problematic when the film and substrate have different crystal structures and lattice constants or when the substrate surface is not atomically smooth. One example of a poor film/substrate interface is shown in Figure 2 for a $Bi_2Se_3$ film grown on a GaAs (001) substrate.[31] These materials have different crystal structures (rhombohedral vs. zinc-blende cubic) and different in-plane lattice constants (4.14 Å vs. 5.65Å), and the thermally-deoxidized substrate surface is rough. Although the $Bi_2Se_3$ film has the correct phase as determined by x-ray diffraction measurements, domains with multiple orientations form, leading to a film with a high density of structural defects and poor electrical performance. In particular, domains oriented in the undesired $(\bar{1}015)$ orientation appear in addition to domains oriented in the desired (0001) orientation. The film also exhibits defects where two grains with different in-plane orientations meet (Figure 2(a)). $Bi_2Se_3$ films grown on atomically-smooth $c$-plane $Al_2O_3$ substrates show similar, though less severe issues. In general, if a vdW film is deposited on a substrate without any specific substrate preparation steps or control over film nucleation and growth processes, the first few layers will typically be of poor quality. In addition to degraded device performance, the poor crystal quality of the first few atomic layers makes it challenging to grow ultra-thin layers of vdW materials, which often have unusual or interesting properties.[32–36] Controlling the film/substrate interface is the first challenge in vdW epitaxy.

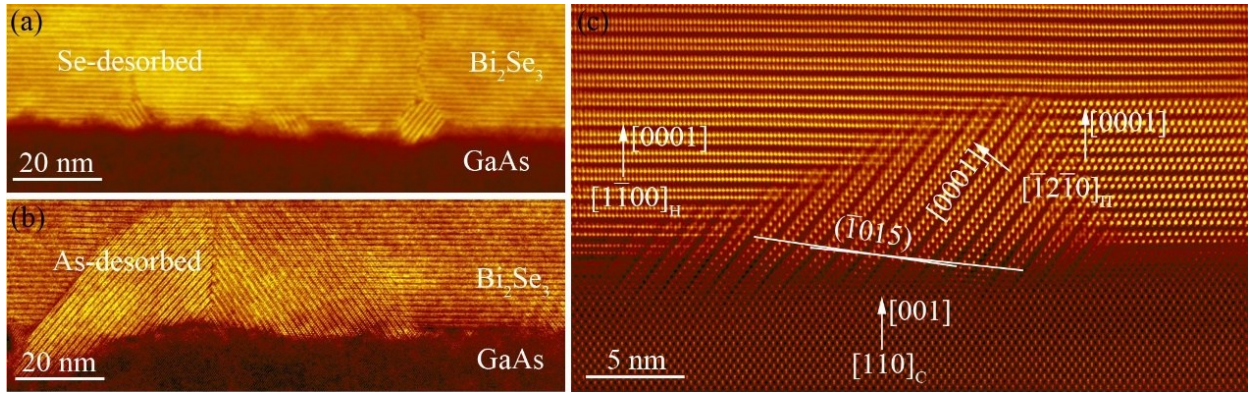

Figure 2 Cross-sectional HAADF-STEM characterization. (a, b) HAADF-STEM images taken along the GaAs $[110]_C$ zone-axis showing the rough interface of $Bi_2Se_3$/GaAs in the Se-desorbed and As-desorbed samples, respectively. (c) Zoomed-in atomic-resolution image of the As-desorbed sample showing $(0001)_H$ oriented growth of $[1100]_H$ and $[\bar{1}2\bar{1}0]_H$ $Bi_2Se_3$ domains on a flat area of the GaAs$(001)_C$ substrate, and $(\bar{1}015)_H$-oriented growth of the $[\bar{1}2\bar{1}0]_H$ $Bi_2Se_3$ domain on a rough multi-faceted area of the GaAs substrate. Adapted with permission from Ref. 31. Copyright 2022 American Chemical Society.

Even if the film is grown on an atomically-smooth substrate with a good interface, other structural defects can arise. A typical atomic force microscope (AFM) image of a 50 nm $Bi_2Se_3$ thin film grown on *c*-plane $Al_2O_3$ is shown in Figure 3. Three types of structural defects are highlighted: twin or antiphase defects, shown in the white dashed box; spiral defects, shown in the gray dotted circles; and pyramidal growth, shown in the black circle. Twin defects can arise in any system where it is equally energetically favorable to nucleate domains in two different directions. For the case of $Bi_2Se_3$ on sapphire, the energy cost to nucleate a right-pointing domain is the same as nucleating a left-pointing domain. In the absence of any substrate treatment or miscut, right-pointing and left-pointing domains nucleate with equal probability. When the domains grow and meet, they are unable to perfectly coalesce due to their different orientations, and a twin defect is formed.[37] Due to the weak interaction between vdW thin films and 3D substrates, a high density of twin defects is often seen in vdW films. In addition to twin defects, stacking faults can occur. Many vdW materials have multiple polymorphs or stacking sequences that have different properties. Due to the similar formation energy among the polymorphs, stacking faults can arise even in vdW homoepitaxy.[38] A stacking fault that comprises a 60º rotation of one layer with respect to the next can lead to a twin defect upon coalescence of that domain with a domain in which all layers have the same orientation. The density of stacking faults and twin defects are therefore often correlated.

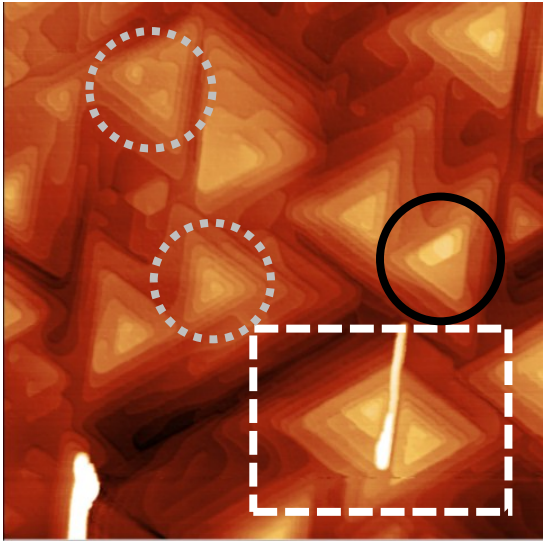

Figure 3 2 μm × 2 μm AFM image of a 50 nm $Bi_2Se_3$ film. Scale bar is 0-20 nm.

In vdW materials, spiral defects can arise when the edge of a domain encounters a step. In most materials, different in-plane directions can grow at different rates. In $Bi_2Se_3$, for example, the $[\bar{1}2\bar{1}]$ direction has twice as many dangling bonds as the $[\bar{2}11]$ direction in equilibrium, so the $[\bar{1}2\bar{1}]$ direction will grow faster.[17] This is why $Bi_2Se_3$ thin films typically comprise triangular domains rather than hexagonal domains. Similar anisotropic growth rates are found in other vdW materials, like InSe along the armchair versus zigzag directions (Figure 4(a)). As the domain grows, it can encounter substrate step edges. If the faster-growing direction encounters the step edge, it will grow over the edge and continue growing smoothly, as shown schematically in Figure 4(b).[22] However, if the domain encounters a step corner, the difference in growth rates between the two different in-plane directions and the difference in $c$-axis lattice constants between the film and substrate can result in spiral growth, as shown schematically in Figure 4(c). Spiral defects can also form when the fast-growing edge of one domain encounters the slow-growing edge of another domain.

Pyramidal growth arises because the energy cost to nucleate a new domain is relatively low, and because many vdW materials do not adequately wet the substrate.[39] If a film is grown on a weakly-interacting substrate, an adatom that has initially incorporated on the substrate at the corner or edge of a domain can diffuse up relatively easily and incorporate on top of the island, since there is no strong bond to the substrate. If this process is repeated, pyramidal growth can result. In addition, due to the difference in chemical composition between the film and substrate, many vdW films do not wet the substrate well, meaning that it is more energetically favorable for adatoms to nucleate on top of existing domains rather than on the substrate. This also promotes pyramidal growth. Additionally, because the in-plane lattice constant of the film and the substrate are often not equal, vdW epitaxy can lead to a high density of registration errors. When the vdW film nucleates on the substrate, the nuclei are not necessarily separated by an integer multiple of the film lattice constant. When these domains meet, they are unable to coalesce smoothly, resulting in regions of compressed or expanded lattice constants.[40]

Finally, the kinetic processes at the growth front can change the predominant growth front and the orientation of the vdW nuclei. Film growth is self-regulated for the vast majority of chalcogenide vdW materials due to the much larger vapor pressure of the chalcogen. At typical growth temperatures, the element with high vapor pressure (e.g. Se) is highly volatile and has a sticking coefficient of zero. The element with low vapor pressure (e.g. Bi), however, has a sticking coefficient of nearly one. If the highly volatile adatoms are not able to interact with and bind to the much less volatile adatom or to substrate atoms, they will simply desorb from the surface and not contribute to growth. These are generally ideal conditions for epitaxy of 3D materials (e.g. GaAs), since the growth rate in these circumstances is limited and determined by the species that is less volatile (e.g. Ga). The highly volatile element (e.g. As) can be oversupplied to ensure the targeted phase formation since excess supply of the volatile species will not significantly influence the growth. For vdW materials, however, the amount of oversupplied volatile atoms has been observed to change nucleation behavior and growth mode extensively.[22,41–44] The driving forces and mechanisms behind this are not yet well understood. For transition metal dichalcogenide MBE growth, an insufficient chalcogen oversupply will cause increased secondary nucleation, high nucleation densities, and small nucleation domain sizes. Increasing the chalcogen flux minimizes secondary nucleation and nucleation domain density and maximizes the lateral size of nucleation domains. However, increasing the chalcogen flux beyond the optimal value can again reduce the film quality.[41,42,44] Similar trends have been observed for the MBE growth of mono- and sesqui-chalcogenides.[45,46] In general, unnecessarily large chalcogen fluxes typically yield pronounced pyramidal growth, columnar island formation, and misoriented domains.

Overall, the difference in lattice constants between the vdW film and the substrate, combined with the weak vdW interaction, poor wetting, and limited understanding of growth kinetics, lead to vdW films with a high density of a variety of defects. The next section will detail a general strategy to mitigate these defects by controlling the nucleation and growth of the vdW film.

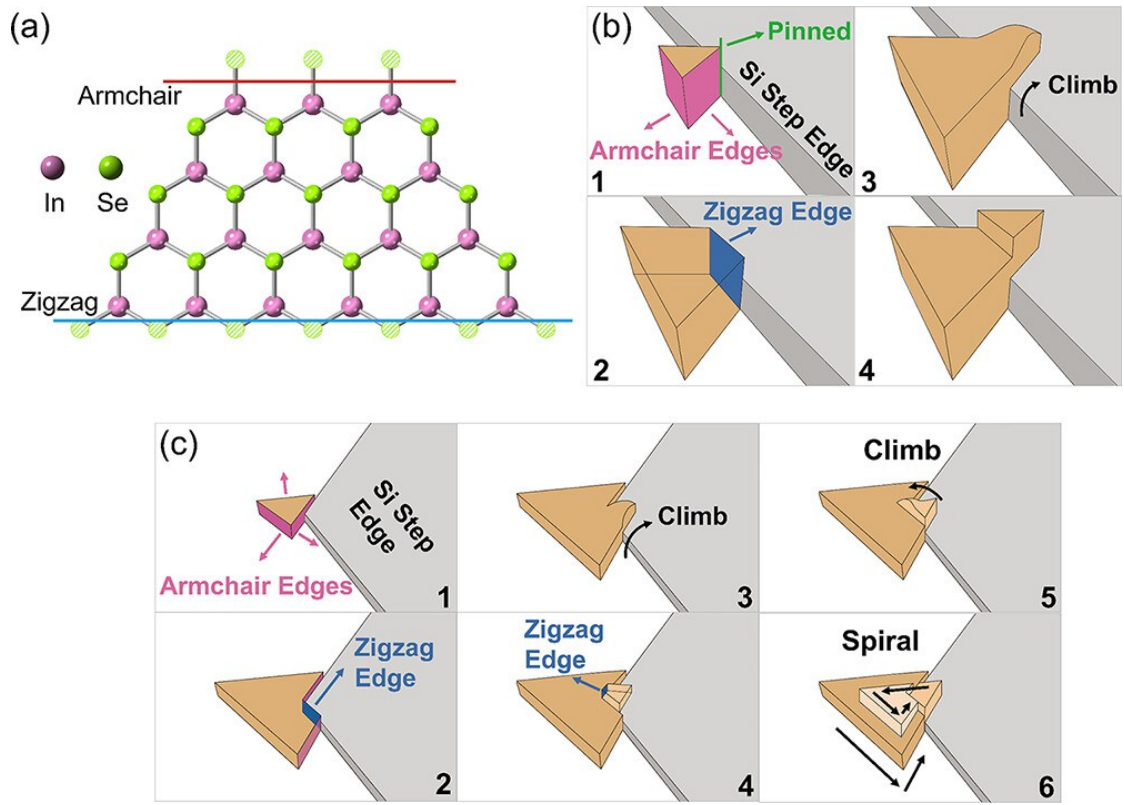

Figure 4 (a) Top view of a single InSe layer flake sketching its atomic arrangement. The pink line intercepts the In armchair edge, and the blue line intercepts the In zigzag edge. (b) Schematic lateral expansion flow of an InSe nucleus climb over a Si(111) step edge. (c) Schematic lateral expansion of an InSe nucleus and spiral growth mode triggered by a step-edge corner. Adapted with permission from Ref. 22. Copyright 2023 American Chemical Society.

## 4. General recipe for the growth of vdW materials by MBE

First and foremost, growth of vdW materials requires a smooth, epi-ready substrate. In conventional MBE, homoepitaxial growth is often used to smooth rough substrates before growing the device layers. However, in vdW epitaxy, it is often impossible to perform homoepitaxial growth in the same MBE chamber due to chemical incompatibility between the film and substrate. Instead, careful attention must be paid to proper substrate treatment prior to growth to minimize rough surfaces and the resulting poor vdW film nucleation. These strategies will depend on the specific substrate of interest. Compound and elemental semiconductor substrates are a common choice due to their excellent electrical and thermal conductivity and ease of semiconductor device integration. However, as they are often un-passivated (i.e., the surface contains a high density of covalent or ionic dangling bonds),[47] these substrates typically have an amorphous oxide layer on the surface, which must be removed prior to growth to generate a flat surface and ensure smooth thin film growth. Wet chemical etching can be used to remove the oxide layer along with other contaminants and passivate the substrate surface (e.g., H-terminated Si (111) with $NH_4F$ etching),[48–50] but these methods are not ideal as the surface is prone to reoxidation in the ambient environment. Instead, an *in situ* thermal anneal in the MBE system is a more common practice to remove the residual oxide from the substrate surface prior to film deposition. Different substrate surface reconstructions such as Si (111) 7×7 and Si (100) 2×2 appear during this process, which is indicative of a clean and atomically smooth surface. The type of surface reconstruction and passivation of Si substrates strongly influences the density of twin domains in the overgrown vdW film.[51,52] In addition, supplying chalcogen fluxes during the UHV anneal is beneficial as it prevents decomposition of compound semiconductor substrates while simultaneously passivating the surface dangling bonds.[53–55] This technique has been widely adapted for various vdW chalcogenide materials grown on III-V substrates semiconductor including GaSe on GaAs, $Bi_2Se_3$ on GaAs, and $Bi_2Te_3$ on InP.[31,56,57] Other chemically-inert substrates such as $Al_2O_3$ and MgO are also commonly used due to their robustness and wide availability. However, these substrates are electrically and thermally insulating, and one cannot rely solely on a UHV anneal alone to obtain a flat surface. Additional *ex situ* treatments such as high temperature (> 1100°C) annealing and/or chemical etching may be needed to reveal atomic step edges at the growth front.[58–60]

With proper substrate treatment processes, atomically smooth surfaces with well-defined step edges can be generated for subsequent vdW layer growth. Fundamentally, vdW epitaxy requires careful control of film nucleation and film growth. Unlike traditional MBE growth, the operator cannot rely on self-assembly alone to produce smooth, high quality films. Instead, separate phases to treat the substrate, nucleate the film, and grow the film are required. A general procedure follows for the growth of layered chalcogenide vdW materials which could be adapted to synthesize non-chalcogenide vdW materials. First, outgas and deoxidize the substrate as needed using the steps discussed above (Fig. 5(a)). **Step 1**: deposit the film on the substrate at a low substrate temperature ($T_{sub}$), ensuring that enough material is deposited to completely cover the substrate with no pinholes; typically, ~5-10 nm (Fig. 5(b)). Appropriate flux ratios should be used to ensure the correct stoichiometry of the film, but the precise growth temperature of this step is not critical so long as all the deposited material sticks to the substrate. **Step 2**: keeping the chalcogen flux supplied but removing the metal flux, heat the substrate to a temperature higher than the thermal decomposition temperature of the film ($T_d$) as determined by reflection high energy electron diffraction (RHEED) analysis (Fig. 5(c)). Anneal the substrate at this temperature until the RHEED pattern matches the bare substrate; typically, 10-30 minutes. Steps 1 and 2 may be repeated multiple times if necessary. **Step 3**: cool the substrate to a temperature 50-100ºC below the thermal decomposition temperature and deposit a few monolayers of the film (Fig. 5(d)). **Step 4**: heat the substrate to near the thermal decomposition point and deposit the rest of the film, ensuring that the deposition rate is higher than the decomposition rate (Fig. 5(e)). The overall growth rate will be lower than expected since the film is growing and re-evaporating at the same time. **Step 5**: cool the substrate in a chalcogen atmosphere until the chalcogen desorption rate from the film is negligible (Fig. 5(f)). Close the chalcogen source and cool the film to room temperature. Cap the film if necessary before removing it from the vacuum chamber.

The purpose of this procedure is to improve wetting of the substrate by the film, reduce screw dislocations, and reduce pyramidal growth. In Step 1, the very low deposition temperature reduces adatom mobility and ensures that the entire substrate is covered by the film regardless of the wettability of the substrate. However, the low deposition temperature also means that film is of poor crystalline quality. Step 2 removes this low-quality film and leaves behind a substrate with a significantly improved wettability. It is likely that the substrate surface has changed in multiple ways. First, some or all of the substrate bonds may be chalcogen-passivated. This is likely the case for substrates that were initially unpassivated as well as for highly reactive substrates.[54,61–63] A chalcogen-passivated substrate is likely to be easier for the film to wet since the metal adatoms can bond to the chalcogen atoms. However, this cannot be the only effect, since merely annealing the substrate in the chalcogen atmosphere does not produce the same improved wettability regardless of the time and temperature of the anneal. It may be the case that, in addition to chalcogen passivation, some nanoscale film nuclei remain on the surface, bonded at step edges or other similar sites with a high density of dangling bonds.[64] These nuclei may help promote the growth of the film in subsequent steps.

In Steps 3 and 4, the high-quality film is grown. The growth of the film at a high temperature enhances adatom mobility and increases grain size. However, it is typically difficult or impossible to nucleate a vdW film near the thermal decomposition point due to the reduced sticking coefficient and the wetting issue. During the low temperature deposition in step 3, film islands nucleate, grow, and coalesce across the substrate. The covalent bonds formed during the process allow the film to remain stable up to the higher temperature of step 4. The high temperature growth significantly improves the structural quality of the film in multiple ways. First, the higher temperature increases adatom mobility, making incorporation at a step edge more likely and reducing pyramidal growth, flattening the film. Second, the density of screw defects is reduced. There are two possible mechanisms for this effect. The additional thermal energy of the high temperature growth may allow both the fast and slow growth directions to grow over step edges, either by changing the number of dangling bonds at the edges of the domains or simply by increasing growth rates. This idea is supported by AFM data showing that films grown at elevated temperatures can have hexagonal rather than triangular domains, indicating that all six edges of the domains are growing with equal rates.[45] In addition, the elevated temperatures may cause re-evaporation of domains that are pinned at step corners. These domains may be less stable than domains that are adjacent to step edges or that have nucleated in the middle of a step, since domains pinned at corners are forced to grow in a non-ideal geometry. Third, the high temperature eliminates small domains and mis-oriented crystallites by enhancing their evaporation compared to well-oriented large domains, which also reduces pyramidal growth and improves the overall structural quality of the film. It can also reduce the density of twin defects if one domain orientation is less energetically favorable than the other.[15] It should be reemphasized that an increased metal flux is required to compensate for the film evaporation at elevated temperatures: the deposition rate must be greater than the re-evaporation rate, otherwise no film will grow.

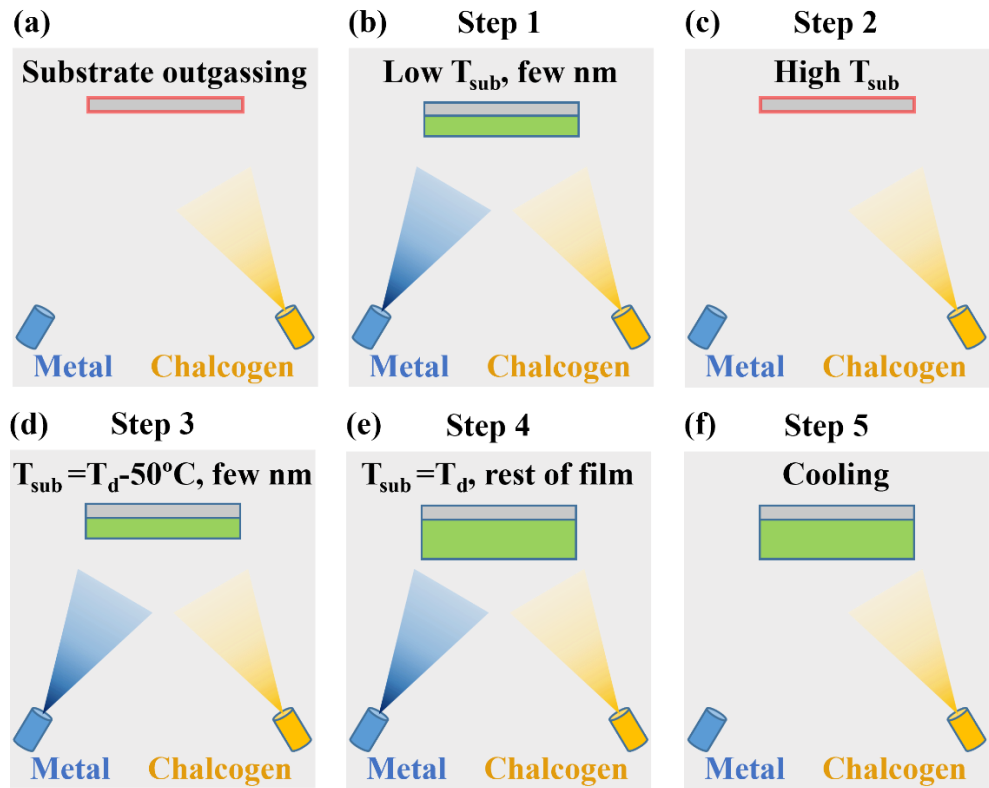

Figure 5 Schematic of vdW film growth process. Substrate is shown in gray, and film is shown in green. $T_{sub}$ is the substrate temperature and $T_d$ is the film decomposition temperature.

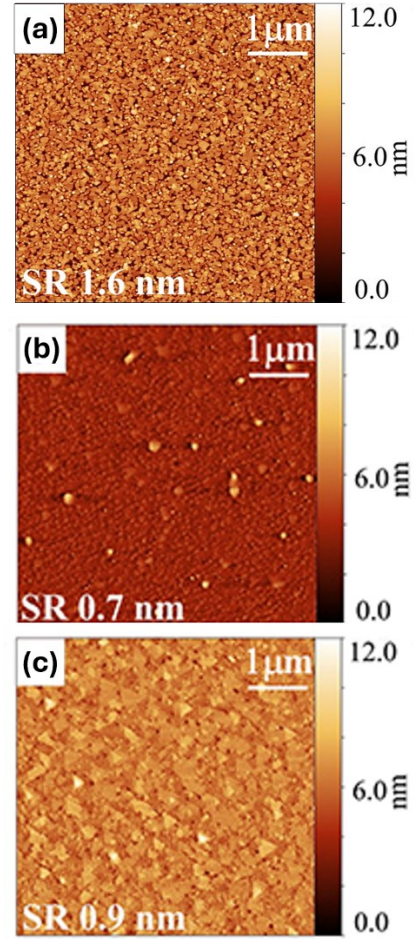

Figure 6 AFM images of 4 nm of $Bi_2Se_3$ deposited on $c$-plane $Al_2O_3$. (a) No substrate pre-treatment; two-temperature deposition. (b) Substrate pre-treatment; one-temperature deposition. (c) Substrate pre-treatment; two-temperature deposition. Reproduced with permission from [32].

The impact of each step of this method can be seen in Figure 6 for a 4 nm thick $Bi_2Se_3$ films grown on $c$-plane $Al_2O_3$ substrates.[32] Figure 6(a) shows an AFM image of a film grown skipping steps 1 and 2: the substrate was not pre-treated, but the film was deposited using two temperature steps. The resulting film comprises many small grains and is not continuous in Hall effect measurements. This is likely due to the fact that $Bi_2Se_3$ does not easily wet $Al_2O_3$. Figure 6(b) shows an AFM image of a film deposited on pretreated substrate (steps 1 and 2) but only grown at one temperature. The film is continuous, but the grain size is very small. Finally, in Figure 6(c), the film is grown using the procedure outlined in Figure 5 and shows well-defined grains despite being only 4 nm thick.

Choosing the temperature and metal flux in step 4 is critical to high-quality film growth. In Figure 7, AFM images of deposition of a GaSe film on a GaAs (111)B substrate are shown.[43] At a substrate temperature of 440ºC, the film decomposes faster than it is deposited, resulting in a roughening of the substrate with no GaSe deposition (Figure 7(a)). At 420ºC, a GaSe film forms, but it has visible cracks due to film decomposition. Reducing the substrate temperature to 375ºC results in a coalesced GaSe film, but the grains are irregularly shaped. However, depositing the film at 420ºC while using a higher gallium flux to compensate for the decomposition results in a film with clear, triangular domains and a reduced density of spiral growth areas. This film clearly shows that growing at the highest possible substrate temperature while maintaining a sufficient metal flux to overcome film decomposition leads to films with a desirable morphology. The only downside to this method is that it can be difficult to deposit films with an accurate thickness due to the competing effects of deposition and decomposition.

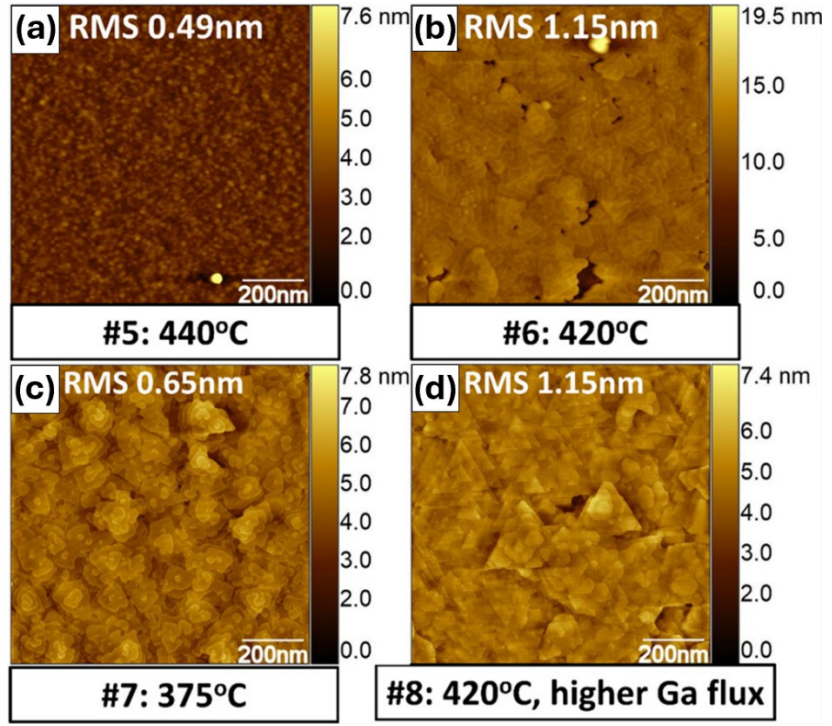

Figure 7 AFM images of GaSe grown on GaAs(111)B substrates. At 440°C, no film is deposited (a). At 420°C, a film grows with visible cracks (b). At 375°C, the film grains show an irregular shape (c). At 420°C with a higher gallium flux, the film shows well-formed triangular domains with few screw defects (d). Reproduced with permission from [43].

In addition to growing high-quality thick films, the multi-step deposition process can also be used to grow ultrathin films. Using traditional co-deposition methods, it can be challenging to synthesize coalesced ultrathin films of most 2D materials on most 3D substrates due to low wettability of the vdW material. For example, 3-layer GaSe films grown on $Al_2O_3$ substrates using traditional direct growth results in a cloudy RHEED pattern and an AFM image comprising droplets, as shown in Figure 8(a, b).[36] This is caused by the poor wettability of GaSe on sapphire. If the deposition were to continue, the droplets would eventually be buried and incorporated into the film. However, using the multi-step growth method results in a streaky RHEED pattern and a coalesced film after only 3 layers of GaSe, as shown in Figure 8(c, d). This again shows the power of this method to produce high-quality vdW thin films after only a few nanometers. This general process works for a wide range of 2D materials grown on various types of substrates.[15,32,43,65] Overall, this recipe enables separate control of the nucleation, coalescence, and growth phases, resulting in high-quality thin films of 2D materials.

## 5. Conclusion

The multi-step process discussed above can be used to mitigate structural defects in vdW films grown on non-vdW substrates, including twin defects, spiral growth, and pyramidal growth. There are multiple reports of improvements in film quality using parts of this recipe, e.g. two-step growth only, substrate preparation only, or high temperature growth only.[66,66–69] However, combining all of these techniques into one recipe reproducibly produces high-quality vdW thin films on a range of substrates. Although this article focused on MBE growth, these techniques may be applicable to other synthesis methods including (metal organic) chemical vapor deposition, pulsed laser deposition, or sputtering. The ability to scalably grow vdW thin films and heterostructures with a reduced defect density is important for a wide range of applications including next-generation optoelectronics, quantum devices, and fundamental physics studies.

## Acknowledgements

This research was conducted at the Pennsylvania State University Two-Dimensional Crystal Consortium – Materials Innovation Platform which is supported by NSF cooperative agreement DMR-2039351.

## Data availability

Data underlying the results presented in this paper are available in the original manuscripts.

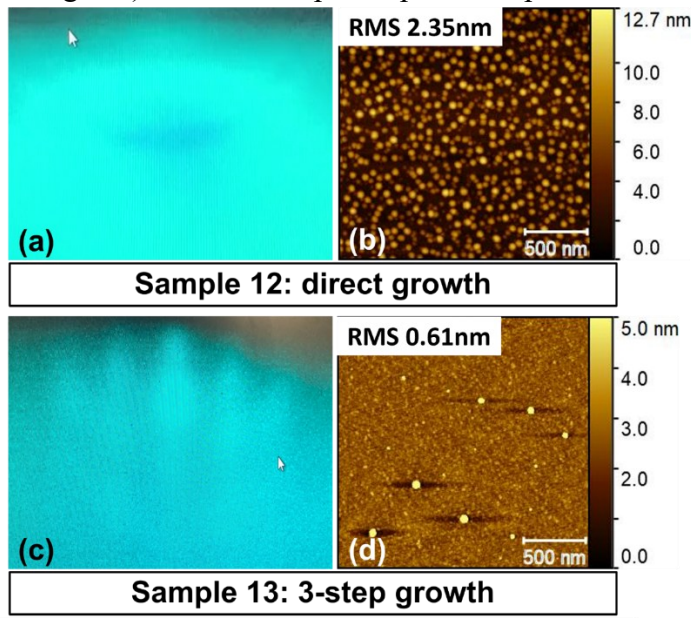

Figure 8 RHEED (a, c) and AFM (b, d) images of 3-layer GaSe grown on sapphire using direct growth (a, b) and the multi-step growth recipe described above (c, d). Reproduced with permission from [36].

## References

[1] Y. Liu, N. O. Weiss, X. Duan, H.-C. Cheng, Y. Huang, X. Duan, *Nat. Rev. Mater.* **2016**, *1*, 1.
[2] A. K. Geim, I. V. Grigorieva, *Nature* **2013**, *499*, 419.
[3] A. Castellanos-Gomez, X. Duan, Z. Fei, H. R. Gutierrez, Y. Huang, X. Huang, J. Quereda, Q. Qian, E. Sutter, P. Sutter, *Nat. Rev. Methods Primer* **2022**, *2*, 1.
[4] K. S. Novoselov, A. Mishchenko, A. Carvalho, A. H. Castro Neto, *Science* **2016**, *353*, aac9439.
[5] L. A. Walsh, C. L. Hinkle, *Appl. Mater. Today* **2017**, *9*, 504.
[6] B. A. Joyce, *Rep. Prog. Phys.* **1985**, *48*, 1637.
[7] A. Y. Cho, *J. Cryst. Growth* **1991**, *111*, 1.
[8] M. Henini, *Molecular Beam Epitaxy From Research to Mass Production*, Elsevier Science, **2013**.
[9] F. Arciprete, J. E. Boschker, S. Cecchi, E. Zallo, V. Bragaglia, R. Calarco, *Adv. Mater. Interfaces* **2022**, *9*, 2101556.
[10] A. Koma, *Thin Solid Films* **1992**, *216*, 72.
[11] W. Mortelmans, S. De Gendt, M. Heyns, C. Merckling, *Appl. Mater. Today* **2021**, *22*, 100975.
[12] J. Dong, L. Zhang, X. Dai, F. Ding, *Nat. Commun.* **2020**, *11*, 5862.
[13] A. Koma, *J. Cryst. Growth* **1999**, *201–202*, 236.
[14] A. K. Katiyar, A. T. Hoang, D. Xu, J. Hong, B. J. Kim, S. Ji, J.-H. Ahn, *Chem. Rev.* **2024**, *124*, 318.
[15] Z. Wang, S. Law, *Cryst. Growth Des.* **2021**, *21*, 6752.
[16] Y. Wang, T. P. Ginley, S. Law, *J. Vac. Sci. Technol. B* **2018**, *36*, 02D101.
[17] Y. Liu, M. Weinert, L. Li, *Phys. Rev. Lett.* **2012**, *108*, 115501.
[18] C. Liu, T. Liu, Z. Zhang, Z. Sun, G. Zhang, E. Wang, K. Liu, *Nat. Nanotechnol.* **2024**, *19*, 907.
[19] Y. Nie, A. T. Barton, R. Addou, Y. Zheng, L. A. Walsh, S. M. Eichfeld, R. Yue, C. R. Cormier, C. Zhang, Q. Wang, C. Liang, J. A. Robinson, M. Kim, W. Vandenberghe, L. Colombo, P.-R. Cha, R. M. Wallace, C. L. Hinkle, K. Cho, *Nanoscale* **2018**, *10*, 15023.

[20] M. Hilse, J. Rodriguez, J. Gray, J. Yao, S. Ding, D. S. H. Liu, M. Li, J. Young, Y. Liu, R. Engel-Herbert, **2024**, DOI 10.48550/arXiv.2404.12578.
[21] M. Hilse, N. Trainor, A. R. Graves, R. Xiao, M. Stanley, Y. Ou, D. S. H. Liu, R. Engel-Herbert, A. Richardella, S. Law, J. M. Redwing, in *Compr. Semicond. Sci. Technol. Second Ed.* (Ed.: R. Fornari), Elsevier, Oxford, **2025**, pp. 329–375.
[22] D. S. H. Liu, M. Hilse, A. R. Lupini, J. M. Redwing, R. Engel-Herbert, *ACS Appl. Nano Mater.* **2023**, *6*, 15029.
[23] H. Kim, J. Bae, S. J. Pearton, F. Ren, J. Kim, G.-H. Lee, *2D Mater.* **2025**, *12*, 022003.
[24] D.-H. Yoon, K.-S. Lee, J.-B. Yoo, T.-Y. Seong, *Jpn. J. Appl. Phys.* **2002**, *41*, 1253.
[25] H. Zhao, X. Li, Y. Zhao, M. Tan, W. Yang, T. Wei, S. Lu, *J. Cryst. Growth* **2024**, *632*, 127632.
[26] Y. Bogumilowicz, J. M. Hartmann, M. Martin, A. M. Papon, D. Muyard, Z. Saghi, S. David, T. Baron, *J. Cryst. Growth* **2025**, *667*, 128235.
[27] M. A. Herman, H. Sitter, *Molecular Beam Epitaxy: Fundamentals and Current Status*, Springer, Berlin, Heidelberg, **1996**.
[28] A. Y. Cho, J. R. Arthur, *Prog. Solid State Chem.* **1975**, *10*, 157.
[29] S. Franchi, in *Mol. Beam Epitaxy* (Ed.: M. Henini), Elsevier, Oxford, **2013**, pp. 1–46.
[30] W. K. Burton, N. Cabrera, F. C. Frank, N. F. Mott, *Philos. Trans. R. Soc. Lond. Ser. Math. Phys. Sci.* **1951**, *243*, 299.
[31] Y. Liu, W. Acuna, H. Zhang, D. Q. Ho, R. Hu, Z. Wang, A. Janotti, G. Bryant, A. V. Davydov, J. M. O. Zide, S. Law, *ACS Appl. Mater. Interfaces* **2022**, *14*, 42683.
[32] S. Nasir, W. J. Smith, T. E. Beechem, S. Law, *J. Vac. Sci. Technol. A* **2023**, *41*, 012202.
[33] S. Borisova, J. Krumrain, M. Luysberg, G. Mussler, D. Grützmacher, *Cryst. Growth Des.* **2012**, *12*, 6098.
[34] M. Eddrief, F. Vidal, B. Gallas, *J. Phys. Appl. Phys.* **2016**, *49*, DOI 10.1088/0022-3727/49/50/505304.
[35] S. S. Hong, W. Kundhikanjana, J. J. Cha, K. Lai, D. Kong, S. Meister, M. A. Kelly, Z. X. Shen, Y. Cui, *Nano Lett.* **2010**, *10*, 3118.
[36] M. Yu, L. Murray, M. Doty, S. Law, *J. Vac. Sci. Technol. A* **2023**, *41*, 032704.
[37] X. Zou, Y. Liu, B. I. Yakobson, *Nano Lett.* **2013**, *13*, 253.
[38] W. Mortelmans, A. Nalin Mehta, Y. Balaji, S. El Kazzi, S. Sergeant, M. Houssa, S. De Gendt, M. Heyns, C. Merckling, *2D Mater.* **2020**, *7*, 025027.
[39] B. Lü, G. A. Almyras, V. Gervilla, J. E. Greene, K. Sarakinos, *Phys. Rev. Mater.* **2018**, *2*, 063401.
[40] M. Chubarov, T. H. Choudhury, D. R. Hickey, S. Bachu, T. Zhang, A. Sebastian, A. Bansal, H. Zhu, N. Trainor, S. Das, M. Terrones, N. Alem, J. M. Redwing, *ACS Nano* **2021**, *15*, 2532.
[41] P. M. Litwin, M. G. Sales, V. Nilsson, P. V. Balachandran, C. Constantin, S. McDonnell, in *Low-Dimens. Mater. Devices 2019*, SPIE, **2019**, pp. 40–52.
[42] A. Rajan, K. Underwood, F. Mazzola, P. D. C. King, *Phys. Rev. Mater.* **2020**, *4*, 014003.
[43] M. Yu, S. A. Iddawela, J. Wang, M. Hilse, J. L. Thompson, D. Reifsnyder Hickey, S. B. Sinnott, S. Law, *ACS Nano* **2024**, *18*, 17185.
[44] R. Yue, Y. Nie, L. A. Walsh, R. Addou, C. Liang, N. Lu, A. T. Barton, H. Zhu, Z. Che, D. Barrera, L. Cheng, P.-R. Cha, Y. J. Chabal, J. W. P. Hsu, J. Kim, M. J. Kim, L. Colombo, R. M. Wallace, K. Cho, C. L. Hinkle, *2D Mater.* **2017**, *4*, 045019.
[45] T. P. Ginley, S. Law, *J Vac Sci Technol A* **2021**, *39*, 33401.
[46] T. P. Ginley, Y. Zhang, C. Ni, S. Law, *J. Vac. Sci. Technol. A* **2020**, *38*, 023404.
[47] A. Ohtake, Y. Sakuma, *Appl. Phys. Lett.* **2019**, *114*, 053106.
[48] H. Hirayama, T. Tatsumi, *Appl. Phys. Lett.* **1989**, *54*, 1561.
[49] S. S. Iyer, M. Arienzo, E. de Frésart, *Appl. Phys. Lett.* **1990**, *57*, 893.
[50] S. -K. Yang, S. Peter, C. G. Takoudis, *J. Appl. Phys.* **1994**, *76*, 4107.
[51] J. E. Boschker, J. Momand, V. Bragaglia, R. Wang, K. Perumal, A. Giussani, B. J. Kooi, H. Riechert, R. Calarco, *Nano Lett.* **2014**, *14*, 3534.
[52] J. Momand, J. E. Boschker, R. Wang, R. Calarco, B. J. Kooi, *CrystEngComm* **2018**, *20*, 340.

[53] A. Ohtake, S. Goto, J. Nakamura, *Sci. Rep.* **2018**, *8*, 1220.
[54] M. Yu, J. Wang, S. A. Iddawela, M. McDonough, J. L. Thompson, S. B. Sinnott, D. Reifsnyder Hickey, S. Law, *J. Vac. Sci. Technol. B* **2024**, *42*, 033201.
[55] Q. Zhang, M. Hilse, W. Auker, J. Gray, S. Law, *ACS Appl. Mater. Interfaces* **2024**, *16*, 48598.
[56] J. S. Lee, A. Richardella, D. W. Rench, R. D. Fraleigh, T. C. Flanagan, J. A. Borchers, J. Tao, N. Samarth, *Phys. Rev. B* **2014**, *89*, 174425.
[57] M. Budiman, A. Yamada, M. Konagai, *Jpn. J. Appl. Phys.* **1998**, *37*, 4092.
[58] R. Trice, M. Yu, A. Richardella, M. Hilse, S. Law, *J. Vac. Sci. Technol. A* **2025**, *43*, 033407.
[59] W. Mortelmans, S. El Kazzi, A. Nalin Mehta, D. Vanhaeren, T. Conard, J. Meersschaut, T. Nuytten, S. De Gendt, M. Heyns, C. Merckling, *Nanotechnology* **2019**, *30*, 465601.
[60] H. Zhu, N. Nayir, T. H. Choudhury, A. Bansal, B. Huet, K. Zhang, A. A. Puretzky, S. Bachu, K. York, T. V. Mc Knight, N. Trainor, A. Oberoi, K. Wang, S. Das, R. A. Makin, S. M. Durbin, S. Huang, N. Alem, V. H. Crespi, A. C. T. van Duin, J. M. Redwing, *Nat. Nanotechnol.* **2023**, *18*, 1295.
[61] C. González, I. Benito, J. Ortega, L. Jurczyszyn, J. M. Blanco, R. Pérez, F. Flores, T. U. Kampen, D. R. T. Zahn, W. Braun, *J. Phys. Condens. Matter* **2004**, *16*, 2187.
[62] F. S. Turco, C. J. Sandroff, M. S. Hedge, M. C. Tamargo, *J. Vac. Sci. Technol. B Microelectron. Process. Phenom.* **1990**, *8*, 856.
[63] K. S. Wickramasinghe, C. Forrester, M. R. McCartney, D. J. Smith, M. C. Tamargo, **2024**, DOI 10.48550/arXiv.2405.09371.
[64] S. Chegwidden, Z. Dai, M. A. Olmstead, *J. Vac. Sci. Technol. A* **1998**, *16*, 2376.
[65] Q. Zhang, K. Wang, W. Auker, M. Hilse, S. Law, *Cryst. Growth Des.* **2025**, *25*, 2476.
[66] S. E. Harrison, S. Li, Y. Huo, B. Zhou, Y. L. Chen, J. S. Harris, *Appl. Phys. Lett.* **2013**, *102*, 171906.
[67] C. Vergnaud, M.-T. Dau, B. Grévin, C. Licitra, A. Marty, H. Okuno, M. Jamet, *Nanotechnology* **2020**, *31*, 255602.
[68] K. Ueda, Y. Hadate, K. Suzuki, H. Asano, *Thin Solid Films* **2020**, *713*, 138361.
[69] I. Kim, J. Ryu, E. Lee, S. Lee, S. Lee, W. Suh, J. Lee, M. Kim, H. seok Oh, G.-C. Yi, *NPG Asia Mater.* **2024**, *16*, 1.